\RequirePackage{fix-cm}

\documentclass[twocolumn,epjc3]{svjour3}

\RequirePackage[T1]{fontenc}

\smartqed  % flush right qed marks, e.g. at end of proof

\RequirePackage{amsfonts}
\RequirePackage{graphicx}
\RequirePackage{mathptmx}      % use Times fonts if available on your TeX system
\RequirePackage{flushend}
\RequirePackage[numbers,sort&compress]{natbib}
\RequirePackage[colorlinks,citecolor=blue,urlcolor=blue,linkcolor=blue]{hyperref}

\newcommand{\R}{R\'enyi }
\journalname{Eur. Phys. J. C}

\RequirePackage{psfrag}
\RequirePackage{textcomp}
\RequirePackage{amsmath,amssymb,color}

\begin{document}

\title{Hawking-\R black hole thermodynamics, Kiselev solution, and cosmic censorship}
\author{Viktor G. Czinner\thanksref{e1,addr1} 
	\and 
	Hideo Iguchi\thanksref{e2,addr2}}
	
\thankstext{e1}{e-mail: vczinner@gmail.com}
\thankstext{e2}{e-mail: iguchi.hideo@nihon-u.ac.jp (corresponding author)}

\institute{Pa\c{c}o de Arcos, Portugal \label{addr1} 
	\and
	Laboratory of Physics, College of Science and Technology, Nihon University, 274-8501 Narashinodai, Funabashi, Chiba, Japan \label{addr2}}
%\date{\today}
\date{Received: date / Accepted: date}

\maketitle

\begin{abstract}
Explicit example, where the Hawking temperature of a black hole horizon is compatible with the black hole's \R entropy thermodynamic description, is cons\-truct\-ed. It is shown that for every static, spherically symmetric, vacuum black hole space-time, a corresponding black hole solution can be derived, where the Hawking temperature is identical with the \R temperature, i.e.~the one obtained from the \R entropy of the black hole via the {\it 1st law} of thermodynamics. In order to have this Hawking-\R type thermodynamic property, the black holes must be sur\-round\-ed by an anisotropic fluid in the form of a Kiselev metric, where the properties of the fluid are uniquely determined by the mass of the black hole, $M$, and the \R parameter, $\lambda$. In the simplest Schwarzschild scenario, the system is found to be thermodynamically unstable, and the {\it 3rd law} of thermodynamics seems to play the role of a cosmic censor via placing an upper bound on the black hole's mass, by which preventing the black hole from loosing its horizon(s).
\end{abstract}


\section{Introduction}

The Bekenstein--Hawking entropy of black holes \cite{B,Betal,H,H2} follows the horizon area law, $S_{BH}=\pi r_h^2$, which is well known to be nonadditive, resulting the long standing problem of thermodynamically unstable Schwarz\-schild black holes with negative heat capacity. The standard thermodynamic approach is
%It's path integral derivation \cite{GH},
based on the Boltzmann-Gibbs type, $S_{BG}= -\sum p_i\ln p_i$, statistical entropy formula, which may not be the most appropriate choice to describe strongly gravitating systems, such as a black hole.
%has some unresolved issues as well \cite{Wald,CI2},
A possible way of addressing this problem is to consider more general entropy definitions. Nonextensive thermodynamics \cite{Tsallis}, such as the one based on the Tsallis entropy \cite{Tsallis2}, $S_T= -\tfrac{1}{\lambda}\sum(p_i^{1-\lambda}-p_i)$, aims at describing dynamically and statistically entangled systems, and long-range type interactions. Gravitation, black holes in particular, are eloquent examples for such a behaviour \cite{bhnonext}, where the Tsallis model \cite{TC} may serve as a more reasonable alternative to the Boltzmann-Gibbs description. The parameter value $\lambda$, in the definition of the Tsallis formula, generally arises from the concrete, specific problems in question, e.g.~from quantum entanglement phenomena \cite{qec}, finite size reservoir effects \cite{fr}, or other long-range type interaction effects \cite{lr}. In the limit when $\lambda\rightarrow 0$, one can show that $S_T \rightarrow S_{BG}$.

In contrast to the Boltzmann-Gibbs description, the Tsallis entropy has the well known property of being nonadditive for composition, $S_T(A,B)=S_T(A)+S_T(B)+\lambda S_T(A)S_T(B)$. This feature, on the one hand, naturally represents the entanglement property of the system in question, on the other hand, it results some {\it 0th law} compatibility problems when one wants to describe the system with a consistent thermodynamics, equipped with a well defined temperature function \cite{Abe,BV}.

Another known, parametric extension of the Boltzmann-Gibbs entropy is the \R entropy formula \cite{Renyi,Renyibook}, $S_R= -\tfrac{1}{\lambda}\ln\sum p_i^{1-\lambda}$, which is related to the
Tsallis one as $S_R = \frac{1}{\lambda}\ln[1+\lambda S_T]$. Compared to the Tsallis entropy however,
it has the important property of being additive for composition, which can result a consistent thermodynamical description with a {\it 0th law} compatible temperature function, while also accounting for the entangled nature of the system via the $\lambda$ parameter. As a consequence, the \R entropy is more suitable for a consistent physical interpretation to the black
hole thermodynamic problem in general. In the limit when $\lambda\rightarrow 0$, it also recovers the Boltzmann-Gibbs formula.

In our earlier works \cite{BC,CI,CI2}, we studied the \R model of black hole thermodynamics by considering the
\begin{equation}\label{SR}
 S_R = \frac{1}{\lambda}\ln[1+\lambda S_{BH}]\ ,
\end{equation}
expression for the thermodynamic entropy. We proposed that black holes may be better described by a Tsallis-type statistical entropy rather than the standard Boltzmann-Gibbs model, and for a consistent thermodynamical description, we considered its "formal logarithm", the \R entropy. As a result, we showed that the \R thermodynamic behaviour of Sch\-warz\-schild and Kerr black holes are very similar to the original Boltzmann-Gibbs one on anti-de Sitter backgrounds \cite{HP}. By this analogy, the $\lambda$ parameter of the \R model has been connected to the cosmological constant via identical forms of the temperature function, and the stability properties of the two descriptions were also shown to be similar, exhibiting critical phenomena like Hawking--Page phase transition \cite{HP}, and therefore resulting thermodynamically stable, asymptotically flat black holes. Later on, this approach has been followed by numerous authors in investigating various black hole space-times \cite{promsiri2020thermodynamics, tannukij2020thermodynamics,nakarachinda2021effective,promsiri2021solid,abreu2021nature,barzi2022renyi,el2022critical,promsiri2022emergent,wang2023thermodynamics,barzi2023some,tong2024topology}, and also considered in cosmological applications \cite{CM,cc1,cc2,cc3}.

In general, in the \R entropy model, the Hawking temperature of the black hole horizon is different from the thermodynamic \R temperature, and only the \R version exhibits the AdS similarity. In this paper however, we show that in certain cases, the Hawking result can be compatible with the \R description, that is, the two temperatures are identical. We start with the most general, static, spherically symmetric, vacuum black hole solution in standard four dimensional general relativity, with a metric function, $f(r)$, for which the two temperatures are different. We then pose the question whether there exist a modified form of the metric, $\bar f(r,\lambda)$, for which the black hole entropy has the \R form (\ref{SR}), and the thermodynamic \R temperature is identical with the Hawking temperature of the black hole horizon. We search $\bar f(r,\lambda)$ in a form which provides a consistent asymptotic behavior as $\lambda\rightarrow 0$: $\bar f(r,\lambda)\rightarrow f(r)$. From $\bar f(r,\lambda)$, in general, one would not expect to be a known solution of the Einstein equations, however, as it turns out, the obtained result has this property, and we don't have to solve the field equations for its interpretation. The new solution behaves as we were adding a specific anisotropic fluid matter to the original black hole space-time in the form of a Kiselev metric \cite{Kiselev}.

In the following sections, we derive the fluid black hole solution, present its basic thermodynamic properties, and show, that in this system, the {\it 3rd law} of thermodynamics imposes an upper bound on the black hole mass, by which it plays the role of a cosmic censor, preventing the black hole from loosing its horizon(s). Throughout this paper, we work in dimensionless units such that $c = G = \hbar = k_B =1$, and the \R entropy parameter $\lambda$ is also dimensionless.

%All over the paper we use units such as $c=G=k_B=1$.
%All over this paper energy, mass and momentum is measured in multiples of the Planck mass, $M_P$,
%while length and time in multiples of the Planck length, $L_P$. All equations relating
%unlike quantities are to be supported by corresponding powers of $M_P$ and $L_P$.
%The entropy is measured in units of the Boltzmann constant, $k_B$, and the \R entropy parameter $\lambda$ is dimensionless.
%Furtermore we choose the $c=G=k_B=1$ unit system, where one can expresses the Planck constant as $\hbar = L_P M_P$.

\section{Spherical, static, vacuum black holes}

The line element of a black hole can be written in the general form
\begin{equation}
 ds^2 = - f(r) dt^2 + \frac{dr^2}{f(r)} + r^2 d\Omega^2\ ,
\end{equation}
with
\begin{equation}
 f(r) = 1 - \frac{2 M}{r} + f_0(r)\ ,
\end{equation}
where $f_0(r)$ represents the terms that are not mass dependent. For instance, in the case of the Reissner-Nordstr\"om -- de Sitter solution
\begin{equation}\label{rnds}
 f_0(r) = \frac{e^2}{r^2} - \Lambda r^2\ ,
\end{equation}
where $e$ is the electric charge and $\Lambda$ is the cosmological constant. Since we want the new metric to provide a horizon temperature that is equal to the \R temperature, clearly, the modifying term has to be energy dependent, i.e.~it has to depend on the mass of the black hole. In addition, in the $\lambda\rightarrow 0$ limit, we want the original metric recovered, hence we may start looking for a solution in the form
\begin{equation}\label{fbarh}
\bar f(r,\lambda) = 1 - \frac{2 M(1+\lambda h(r))}{r} + f_0(r)\ ,
\end{equation}
where $h(r)$ depends on the radial coordinate only, and in the limit of $\lambda\rightarrow 0$, we get back the original metric. This simple ansatz may seem very restrictive, however we will show that it is in fact very general, which is a consequence of the specific form of the \R entropy of black holes. Let us now rewrite (\ref{fbarh}) as
\begin{equation}\label{fbarg}
\bar f(r,\lambda) = 1 - 2 M g(r,\lambda) + f_0(r)\ ,
\end{equation}
with
\begin{equation}\label{g}
 g(r,\lambda) = \frac{1+\lambda h(r)}{r}\ .
\end{equation}
From the horizon condition, $\bar f(r_h,\lambda)=0$, we have
\begin{equation}\label{M}
 M = \left.\frac{1+f_0(r)}{2g(r,\lambda)}\right|_{r_h}\ ,
\end{equation}
and the horizon- or Hawking temperature is given by
\begin{equation}
 T_H= \left.\frac{\bar f'(r,\lambda)}{4\pi}\right|_{r_h}\ ,
\end{equation}
where prime denotes derivative with respect to $r$. The {\it 1st law} of thermodynamics in the \R entropy description states $dM = T_RdS_R$, and by imposing the condition $T_H \equiv T_R$, we have
\begin{equation}\label{dS}
dS_R = \frac{4\pi}{\left.\bar f'(r,\lambda)\right|_{r_h}} dM\ .
\end{equation}
From (\ref{fbarg}) and (\ref{M}) we have
\begin{equation}\label{fbartilde}
\left.\bar f'\right|_{r_h} = \left[f'_0-2Mg'\right]_{r_h} = \left[f'_0-(1+f_0)\frac{g'}{g}\right]_{r_h}\ ,
\end{equation}
and the derivative of $M$ with respect to $r_h$ results
\begin{equation}\label{Mtilde}
\tilde M = -\frac{1}{2g}\left[\tilde f_0-(1+f_0)\frac{\tilde g}{g}\right]_{r_h}\ ,
\end{equation}
where $\ \tilde{}\ $ denotes derivative with respect to the horizon radius $r_h$.
Since the equality
%\begin{equation}
% \left.\frac{df(r)}{dr}\right|_{r_h} \equiv\  \frac{df(r_h)}{dr_h}\ ,
%\end{equation}
%holds for any function $f$, after short calculation we get
\begin{equation}
 \left.\frac{d\phi(r)}{dr}\right|_{r_h} \equiv\  \frac{d\phi(r_h)}{dr_h}\ ,
\end{equation}
holds for any function $\phi$ that doesn't depend on $M$, after short calculation we get
\begin{equation}
S'_R = \frac{2\pi}{g(r,\lambda)} \ .
\end{equation}
Now using the \R entropy of the black hole as defined in (\ref{SR}), results
%\begin{equation}
% S_R = \frac{1}{\lambda}\ln[1+\lambda\pi r_h^2] \quad\Rightarrow\quad \tilde S(r_h) = \frac{2\pi r_h}{1+%\lambda\pi r_h^2}
%\end{equation}
\begin{equation}\label{g_sol}
 g(r,\lambda) = \frac{2\pi}{S'_R} = \frac{2\pi(1+\lambda S_{BH})}{S'_{BH}} \ ,
\end{equation}
and since we have
\begin{equation}
S_{BH}=\left.\pi r^2\right|_{r_h}\ ,
\end{equation}
(\ref{g_sol}) compared with (\ref{g}) proves that our ansatz is indeed very general, for the solution $h(r)$ we get $h(r)= \pi r^2$, and  the modified metric which satisfies the condition, $T_H\equiv T_R$, takes the form
%\begin{equation}
%\bar f(r,\lambda) = 1 - \frac{2M(1+\lambda\pi r^2)}{r} + f_0(r)\ ,
%\end{equation}
\begin{equation}\label{fbar}
\bar f(r,\lambda) = 1 - \frac{2M}{r} - 2\pi\lambda M r + f_0(r)\ .
\end{equation}

\section{Kiselev black holes with anisotropic fluid}

By applying the {\it 1st law} of thermodynamics with the condition that the \R thermodynamic temperature is identical with the Hawking temperature, the new metric function can be recognized as a known solution of the Einstein equations, namely a Kiselev-type
black hole metric \cite{Kiselev},
\begin{equation}\label{Ksol}
 f_K(r) = 1 - \frac{2M}{r} - \frac{K}{r^{3w+1}} + f_0(r)\ ,
\end{equation}
with the specific parameter values $w=-\frac{2}{3}$ and $K = 2\pi\lambda M$. As shown by Visser \cite{vcqg}, contrary to the common misconception in the literature, the Kiselev black hole space-time is neither perfect fluid, nor is it quintessence in a cosmological sense. With its specific parameter values, our solution describes a black hole surrounded by a fluid matter with energy-momentum tensor
\begin{equation}
 T^{ab} = \text{diag}(\rho, p_r, p_t, p_t) \ ,
\end{equation}
where the energy density, $\rho$, and the anisotropic pressure components $p_r$ and $p_t$ of the fluid are
\begin{equation}
 \rho = -p_r = 2p_t = \frac{\lambda M}{2 r} \ .
\end{equation}
Therefore, all physical parameters of the fluid are uniquely determined by the mass-energy of the black hole, $M$, and the \R parameter, $\lambda$.

By inspecting $\bar f(r,\lambda)$, it is clear that it has some peculiar properties. It describes a space-time whose geometry is asymptotically non flat, and the energy density of the corresponding fluid is linearly coupled to the black hole mass. A physically realistic interpretation of this result may or may not exist, however, an important feature, as shown by Boonserm et al.~\cite{vprd}, is that the energy-momentum tensor of any anisotropic fluid which supports a Kiselev black hole, can always be decomposed (at least formally) into the sum of a perfect fluid component, and either an electromagnetic or a scalar field component. In order to decide between the electromagnetic and the scalar field, the $sign$ of the constant $Kw(1+w)$, constructed from $K$ and $w$ in (\ref{Ksol}), is decisive. If it is negative, the null-energy condition is satisfied, and the energy-momentum tensor of the fluid has an electromagnetic component, instead of a scalar field. In our specific case, $Kw(1+w) = -\frac{4}{9}\pi\lambda M$, and since $\lambda$ is positive, the fluid satisfies the NEC. In addition (for details please refer to \cite{vprd}), the (electrically charged) perfect fluid component of the corresponding energy-momentum tensor has a cosmological constant type equation of state $\rho = -p$, while the anisotropic pressure arise from the electric field.

As a result, the obtained solution may describe (at least formally), a black hole with an electrically charged fluid which is supported by both pressure gradients and its own internally generated electric field. If we also consider that $\bar f(r,\lambda)$ is very general, in the sense that it has a "freely" selectable $f_0(r)$ component, which also can account, for example, for an electric Reissner-Nord\-str\"om part together with a de Sitter part with a cosmological constant, as given in (\ref{rnds}), it may be possible to construct a physically realistic, electrically charged black hole space-time model, surrounded by a cosmological constant type perfect fluid. We leave this question open for future investigations.

The coupling between the black hole mass and the surrounding an\-isotropic fluid is also not unseen phenomena in the literature. For example, in a recent work of Cardoso et al.~\cite{cardoso}, a family of solutions of Einstein’s gravity minimally coupled to an anisotropic fluid has been introduced, describing black holes in order to model the geometry of galaxies harboring supermassive black holes. Later on, this approach has also been considered to model the dark matter problem in galaxies \cite{stuchlik}.

In conclusion, although the metric function, $\bar f(r,\lambda)$, describing a Kiselev black hole with an anisotropic fluid, does not seem to be a physically realistic space-time at first glance, based on the special properties of the corresponding energy-momentum tensor, as well as on the existence of similar, coupled anisotropic fluid approaches for galaxy center black holes and the dark matter problem, it is not impossible that it may have relevance in certain astrophysical problems. In the following section we discuss the thermodynamic aspects of this solution.

\section{Thermodynamics and stability}

\begin{figure}
\noindent\hfil\includegraphics[scale=0.6]{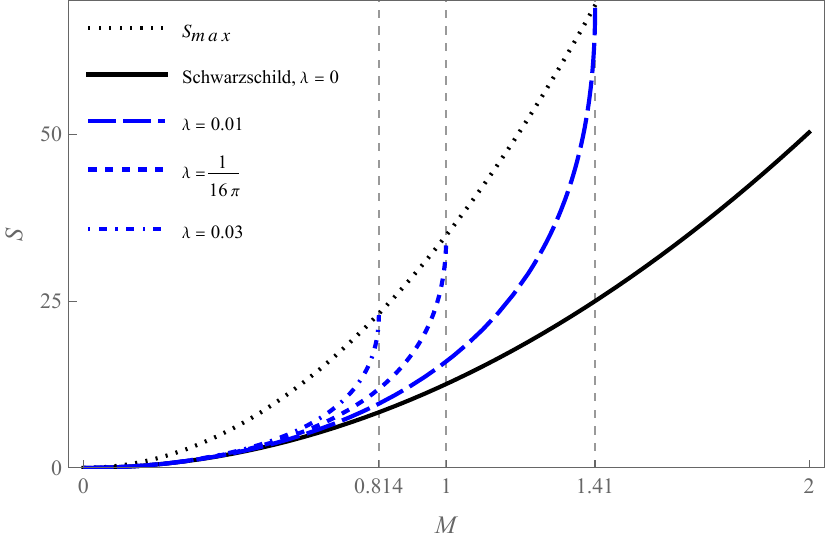}
\caption{\label{figs}The figure shows the \R entropy, $S_R$ of the Kiselev black hole, as a function of $M$ for certain parameter values, $\lambda$. For each nonzero $\lambda$, the entropy approaches its maximum value in the corresponding limit of $M_{max}$, according to Eqs.~(\ref{Mmax}) and (\ref{Smax}).}
\end{figure}

\begin{figure}
\noindent\hfil\includegraphics[scale=0.6]{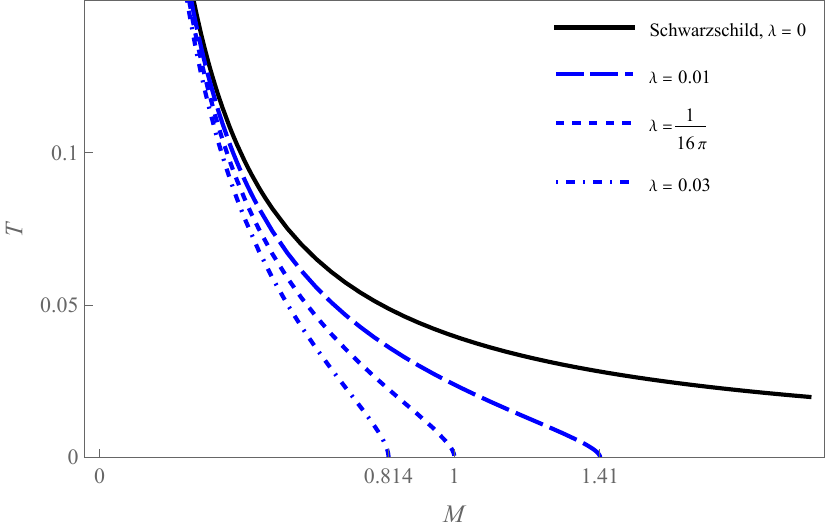}
\caption{\label{figt}The figure shows the temperature, $T_H\equiv T_R$ of the Kiselev black hole, as a function of $M$ for certain $\lambda$ parameter values. For each nonzero $\lambda$, the temperature approaches zero in the corresponding limit of $M_{max}$, according to Eqs.~(\ref{Mmax}) and (\ref{TM}). Since the {\it 3rd law} forbids to reach zero temperature, no naked singularity formation is allowed in the system.}
\end{figure}

The thermodynamic and stability properties of Kiselev black holes (although many of them under the misleading name of black holes with "quint\-essence") have been intensively studied in the literature (see e.g.~\cite{kt1,kt2,kt3,kt4,kt5,kt6,kt7,kt8} and references therein). For simplicity, we only discuss the Schwarz\-schild case here, when $f_0(r)=0$, and only those important aspects which are most relevant within our \R entropy approach. In this scenario, the metric function, $\bar f(r,\lambda)$, can have maximum two horizons \cite{Kiselev}, an inner horizon of black hole type at
\begin{equation}\label{rin}
 r_{in} = \frac{1}{4\pi\lambda M}\left[1-\sqrt{1-16\pi\lambda M^2}\right]\ ,
\end{equation}
and an outer horizon of de Sitter type at
\begin{equation}\label{rout}
 r_{out} = \frac{1}{4\pi\lambda M}\left[1+\sqrt{1-16\pi\lambda M^2}\right]\ .
\end{equation}
Here, we are interested only in the thermodynamics of the black hole type inner horizon.
\begin{figure}
\noindent\hfil\includegraphics[scale=0.6]{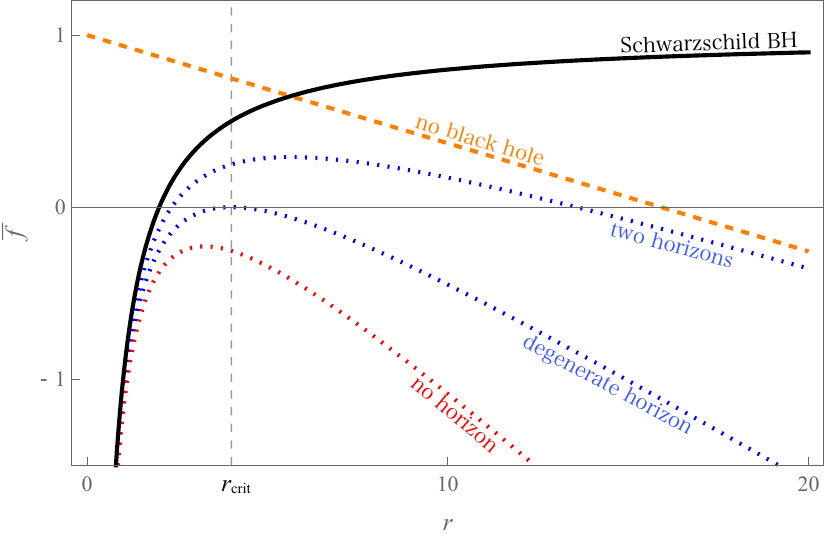}
\caption{\label{figf}The figure shows the possible forms of the metric $\bar f(r,\lambda)$ as a function of $r$. Depending on the metric parameter values $M$ and $\lambda$, the black hole can have two-, one degenerate, or no horizons. The latter case belongs to a naked singularity solution.}
\end{figure}
On Fig.~\ref{figf}, we plotted the possible qualitative behaviors of $\bar f(r,\lambda)$ as a function of the radial coordinate. The solid black curve represents the solution when only the black hole is present without the fluid, while the dashed-orange line denotes the hypothetical case of a pure fluid without the black hole. Both cases have one horizon only, acting as asymptotics from the left and the right. In between, $\bar f(r,\lambda)$ has three possibilities, depending on the discriminant of the equation $\bar f(r,\lambda) = 0$.
If it is positive, we have two horizons as given by (\ref{rin}) and (\ref{rout}), depicted by the blue-dotted curve with two zeros on Fig.~\ref{figf}. In the degenerate case, when the discriminant is zero, the two horizons coincide, represented by the critical blue-dotted curve, which touches the zero line only at the point, $r_{crit}$. In this case we have a maximal black hole mass,
\begin{equation}\label{Mmax}
 M_{max} = \frac{1}{4\sqrt{\pi\lambda}} \ ,
\end{equation}
which we explain shortly. For those cases where the black hole mass would surpass this limit, the $\bar f(r,\lambda)$ function has no zeros, as depicted by the red-dotted curve on Fig.~\ref{figf}, which belongs to the case of a naked singularity. Although, seemingly nothing can prevent the black hole to increase its mass beyond this limit by accreting matter from its surroundings and form a naked singularity, the reason we believe this may not happen is that the black hole mass could only reach this maximal value in the limit where its horizon temperature would go to zero, which is forbidden by the {\it 3rd law} of thermodynamics/black hole mechanics. Indeed, the \R entropy of the black hole, as a function of its mass-energy, reads as
\begin{equation}
 S_R = \frac{1}{\lambda}\ln\left[\frac{2}{1+\sqrt{1-\mu M^2}}\right] \ ,
\end{equation}
with $\mu = 16\pi\lambda$, which we plotted on Fig.~\ref{figs}, for some chosen $\lambda$ parameter values.
It can be seen that the entropy functions reach a maximum value for every $M_{max}$ that corresponds to the chosen $\lambda$ parameter via (\ref{Mmax}) as
\begin{equation}\label{Smax}
S_{max} = 16\pi\ln (2) M_{max}^2 \ ,
\end{equation}
which is $\ln(16)$ times higher than the corresponding pure Schwarz\-schild black hole's entropy. The solid-black curve represents the asymptotic, pure Schwarzschild  solution where $\lambda\rightarrow 0$. For the temperature we have
\begin{equation}\label{Tr}
 T_H \equiv T_R = \left(\frac{dS_R}{dM}\right)^{-1} = \left.\frac{1- \lambda\pi r^2}{4\pi r (1+\lambda\pi r^2)}\right|_{r_h}\ ,
\end{equation}
or expressed as the function of the mass-energy of the black hole
\begin{equation}\label{TM}
T_R(M) = \frac{1-\mu M^2 + \sqrt{1-\mu M^2}}{16\pi M} \ .
\end{equation}
When $M\rightarrow M_{max}$, $T_R\rightarrow 0$, as it can be seen on Fig.~\ref{figt},
where the solid-black curve represents the asymptotic, $T=\tfrac{1}{8\pi M}$, pure Schwarzschild solution again. As we discuss it in details in section \ref{censor}, the {\it 3rd law} of thermodynamics/black hole mechanics states that the zero temperature limit cannot be reached within a finite number of operations, hence it acts as a cosmic censor for this system, and prevents the formation of a naked singularity.

\begin{figure}
\noindent\hfil\includegraphics[scale=0.6]{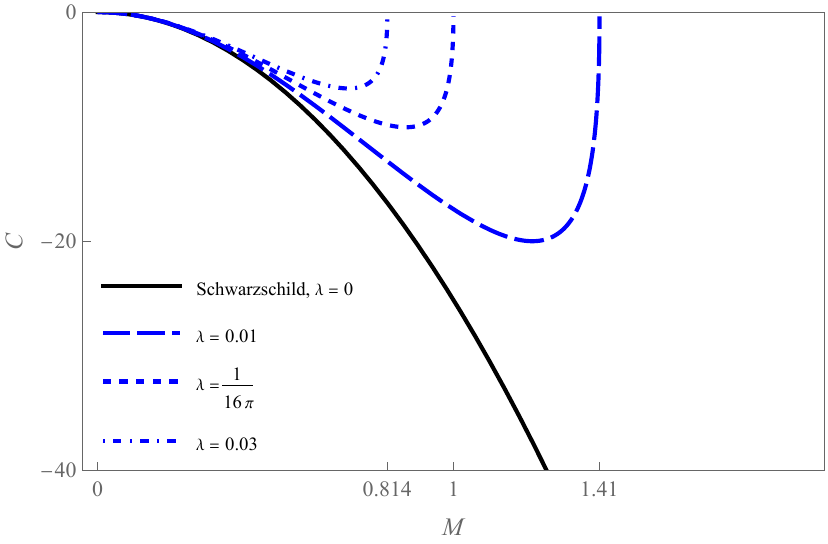}
\caption{\label{figc}The figure shows the heat capacity, $C$, as a function of $M$ for certain parameter values, $\lambda$. At zero $C$ values the system would reach thermodynamic stability, however since these would occure at $M_{max}(\lambda)$, to reach them is forbidden by the {\it 3rd law}, therefore the system is unstable along its lifetime.}
\end{figure}

On Fig.~\ref{figc} we also plotted the corresponding heat capacity curves,
\begin{equation}\label{C}
C(M) = \frac{-S_R'^2(M)}{S_R''(M)} = \frac{-16\pi M^2\sqrt{1-\mu M^2}}{1+(1+\mu M^2)\sqrt{1-\mu M^2}}\ ,
\end{equation}
where prime denotes the derivative with respect to $M$. Since the \R entropy is additive, the sign of the heat capacity is informative about the thermodynamic stability of the black hole. As it can be seen, it is always negative, which means that Schwarzschild-type Kiselev black holes are unstable within this Hawking-\R thermodynamic picture in the presence of the  anisotropic fluid. Without getting into details, one can show by the Poincar\'e stability analysis method \cite{Poincare,CI,CI2}, that the turning points on the heat capacity curves represent a change, from where black holes with higher (or lower) masses become less and less unstable as their mass approaches to $M_{max}$ (or $0$), nevertheless they all remain unstable for any value of $M < M_{max}$.

\section{Cosmic censorship}\label{censor}

The maximal black hole mass criteria, for a given system configuration, can also be interpreted as the "act" of a thermodynamic cosmic censorship \cite{Penrose}. Since $M_{max}$ belongs to the critical point where the horizon temperature would reach the zero value, and from the {\it 3rd law} of thermodynamics/black hole mechanics, we know that: "{\it it is impossible by any physical process to achieve $T=0$} (or alternatively $\kappa=0$), {\it by a finite sequence of operations}", it is interesting to notice that the {\it 3rd law} plays the role of the cosmic censor in this situation. This is because $M_{max}$ also represents the degenerate case, where the inner- and outer horizons would coincide, and by having a larger mass for a given $\lambda$ parameter value, the black hole would loose both of its horizons, and a naked singularity would form. Therefore, there seems to be a very natural thermodynamic reason behind Penrose's cosmic censorship hypothesis in this black hole space-time.

The specific physical mechanisms that can link the {\it 3rd law} to $M_{max}$ is a very interesting question. The maximal mass value is a consequence of the geometry of the Kiselev solution, it designates the limit over which a naked singularity would form. If a black hole is formed with horizons and a mass smaller than $M_{max}$, it can evaporate by Hawking radiation and also accrete matter from its surroundings. If, because of some reasons, the accretion would dominate this process, the black hole mass would grow and the black hole temperature would decrease, according to (\ref{TM}) and Fig.~\ref{figt}. At this point we can invoke the {\it 3rd law} which forbids the black hole to reach $T = 0$ (or alternatively $\kappa=0$) within a finite sequence of operations, therefore it controls the possible maximal accretion limit, and consequently the upper mass limit, $M_{max}$.

The thermodynamic stability properties of the solution are clearly in accordance with this picture. Once an imbalance occurs between the radiation and the accretion process, the black hole will either evaporate completely, or it increases its mass by accretion while gradually cooling down. By cooling down it radiates less, which is a positive feedback and drives the black hole towards the maximal mass value. However, since the {\it 3rd law} forbids this limit where the black hole could reach a thermodynamically stable state according to (\ref{C}) and Fig.~\ref{figc}, the accretion must slow down and stop before reaching $M_{max}$, and the evaporation process could slowly start to kick in again. The black hole's temperature would start to increase and it would produce more and more Hawking radiation. This process should lead to final evaporation eventually, even if the accretion started to dominate again, because after any number of possible oscillations between the accretion and evaporation dominance, the available matter that the black hole could accrete is finite, therefore all of it would be converted into Hawking radiation by the process before the black hole disappears. Either way, the black hole is unstable all along its lifetime, and a naked singularity cannot form if the {\it 3rd law} controls the accretion process in the above manner, hence it acts as a cosmic censor.

Although it is very intriguing to find a connection between the {\it 3rd law} and the cosmic censor hypothesis, identifying a possible concrete physical mechanisms behind it would lead us away from the original objectives of this paper, so we leave this research to a future work.

\section{Discussion and conclusions}

In addition to information theory \cite{it1,it2}, and quantum entanglement phenomena \cite{qe}, where the \R entropy plays a central role in physics, its possible applications to black hole thermodynamics 
\cite{promsiri2020thermodynamics, tannukij2020thermodynamics,nakarachinda2021effective,promsiri2021solid,abreu2021nature,barzi2022renyi,el2022critical,promsiri2022emergent,wang2023thermodynamics,barzi2023some,tong2024topology}, and cosmological problems \cite{CM,cc1,cc2,cc3}, have also been actively investigated lately. Building on our earlier results \cite{BC,CI,CI2}, in this work we considered the problem of describing a black hole with a \R entropy approach by requiring the \R thermodynamic temperature to be identical with the horizon gravity defined Hawking temperature. We showed that the problem is solvable, and constructed a family of explicit examples, where the two temperatures are compatible. Our result is valid for a big class of solutions, namely for all static and spherically symmetric black hole space-times that are coupled to a certain anisotropic fluid matter in the form of a Kiselev-type metric.

The parameters, $w$ and $K$, of the obtained Kiselev solution are uniquely determined by the model, there is no additional freedom left to consider. Only $w=-2/3$ satisfies the condition $T_H \equiv T_R$, and the physical parameters of the corresponding fluid are determined by the black hole's mass, $M$, and the \R parameter, $\lambda$. Although the metric describes an asymptotically non-flat geometry, it has some interesting properties which might make it relevant for certain astrophysical situations. In particular, it may be possible to construct an electrically charged black hole which is embedded into a cosmological constant type dark energy fluid universe, or surrounded by a dark matter type anisotropic fluid, similar to galaxy centre supermassive black holes. Further possible interpretations may also exist.

An interesting question about the model is, why the Hawk\-ing-\R approach singles out the Kiselev metric with the given parameters as the only solution to the problem. More specifically, how does exactly the presence of the anisotropic fluid matter source modify the thermodynamic structure of the black hole in order to result a system, where the thermodynamic entropy function, that is compatible with the Hawking temperature, is the \R entropy. To answer these questions, %however, 
further investigations are required which go beyond the limit of the present work.

To understand more about the solution, we presented the most important thermodynamic properties of the simplest Schwarzschild-type metric, and found that it is generally unstable for any value of the black hole mass. We also showed, that in this system where the fluid parameters are completely determined by $M$ and $\lambda$, there is a maximum value for the black hole mass, above which a naked singularity would form. Since this maximal value corresponds to zero horizon temperature, we point\-ed out that the {\it 3rd law} of thermodynamics/black hole mechanics seems to play the role of a cosmic censor by preventing the black hole to reach this maximal mass limit and loose its horizon(s).

In addition to the Tsallis and R\'enyi descriptions, several other generalized entropies have been proposed lately to study black hole thermodynamics and the holographic dark energy problem. For instance, Nojiri {\it et al. }\cite{Nojiri:2021czz,Nojiri:2021iko,Nojiri:2022aof,Nojiri:2022ljp,Nojiri:2022sfd} conducted intensive research on alternative entropic models, and pointed out the problem of the thermodynamic temperature within these approaches. They even proposed a formula that can treat all entropy models in a unified, multi-parametric manner. In connection with their work, our present result provides an interesting possible direction how to reconcile the Hawking temperature problem with nonstandard entropic models of black hole thermodynamics, or with the holographic dark energy problem in cosmology.

%As a different topic, it has also been suggested that the Kiselev-type black hole metric might be relevant in certain modified theories of gravity \cite{Mannheim1989,Ghosh2016,harada,Burikham2023,Sakti2024}. These models have been studied in various ways to solve the dark energy and dark matter problems in the universe, our result might also have some relevance in these research directions as well.

\begin{acknowledgements}
The authors are grateful for a valuable discussion with T.~Harada, in particular for calling our attention to reference \cite{vcqg}. V.G.Cz.~is thankful for the support of the College of Science and Technology of Nihon University, where most results of this paper have been obtained during a month stay as an Overseas Visiting Fellow in October, 2024.
\end{acknowledgements}

\bibliographystyle{spphys}
\bibliography{references2}

\end{document}